# Neuromorphic Computing with Ferroelectric FinFETs in the Presence of Temperature, Process Variation, Device Aging and Flicker Noise

Sourav De[1], Darsen Lu[1,*], Bo-Han Qiu[1], Md. Aftab Baig[1], Yao-Jen Lee[2,**], Wei-Xuan Bu[1], Cheng-Hsien Tsai[1], Chi-Jen Lin[1], Wei-Chih Tseng[1], Xiao-Shan Huang[1], Changhwan Shin[3], Po-Jung Sung[2], Ta-Chun Cho[2], Chun-Jung Su[2], Chien-Ting Wu[2], Wen-Fa Wu[4], Wen-Kuan Yeh[4]

[1]Institute of Microelectronics, Department of Electrical Engineering, National Cheng Kung University, Taiwan; [2]Taiwan Semiconductor Research Institute, Taiwan; [3]Sungkyunkwan University, South Korea; [4]National Applied Research Laboratories, Taiwan.

email: *darsenlu@mail.ncku.edu.tw, email**: yjlee@narlabs.org.tw , email**: desourav123@gmail.com

## Abstract

This paper reports a comprehensive study on the impacts of temperature-change, process variation, flicker noise and device aging on the inference accuracy of pre-trained all-ferroelectric (FE) FinFET deep neural networks. Multiple-level-cell (MLC) operation with a novel adaptive-program-and-read algorithm with 100ns write pulse has been experimentally demonstrated in 5 nm thick hafnium zirconium oxide (HZO)-based FE-FinFET. With pre-trained neural network (NN) with 97.5% inference accuracy on MNIST dataset as baseline, device to device variation is shown to have negligible impact. Flicker noise characterization at various bias conditions depicts that drain current fluctuation is less than 0.7% with virtually no inference accuracy degradation. The conductance drift of a programmed cell, as an aftermath of temperature change, was captured by a compact model over a wide range of gate biases. Despite significant inference accuracy degradation at 233ºK for a NN trained at 300ºK, gate bias optimization for recovering the accuracy is demonstrated. Endurance above $10^8$ cycles and extrapolated retention above 10 years are shown, which paves the way for edge device artificial intelligence with FE-FinFETs.

## Introduction

Recent advents in research on HZO based FE materials have paved the way for using FE-FinFET for computing in memory (CIM) applications in an attempt to alleviate memory bandwidth limitation induced performance bottleneck in von-Neumann architecture [1]. Pronounced FE in single layer thin film of HZO [2], fast switching, high on/off ratio, excellent linearity, bi-directional operation and good endurance are the key technological factors, which makes FE-FET superior to PCM, RRAM, MRAM and Flash. However, it is well-known that the device to device variations in FE-FETs increases with scaling [3] and the switching becomes stochastic, rather than deterministic. Although recent research has shown 3bit/cell operations in FE-FinFETs [4], which came with additional power consumption, device size and latency, making it awry for large scale applications. Temperature-induced threshold voltage ($V_{th}$) shift in FE-FETs is also found to be very significant in scaled devices [5]. Despite the noise-tolerant nature of DNNs, the CIM architecture, where computational output is determined by the analog current sum, is vulnerable to all types of variation in its basic device building block.

The cornerstone of this work is to mitigate the impact of process, temperature, device aging and flicker noise induced variations in FE-FinFET based CIM operation (Fig. 1) without affecting device scaling, power requirements and latency. The paper begins with a discussion of FE properties in HZO based FE film and fabrication and characterization of deeply scaled FE-FinFETs. Subsequently, we investigate DNN applications focusing on the offline training scenario, where pre-trained weights are programmed to Fe-FinFET devices via a novel closed-loop adaptive-program-and-read algorithm to mitigate the impact of systematic variation on DNN inference accuracy. We characterize and construct statistical compact models for random conductance ($G_{ch}$) variation, temperature effects, flicker noise, and device aging. The model is subsequently applied to a DNN pre-trained with the CIMulator [6] software platform to evaluate inference accuracy degradation. Based on the findings, we propose a new application strategy to minimize the impacts of device non-ideal effects.

## Device Fabrication and Characterization

We fabricate HZO based FE capacitors with an area of $10^4$ μm$^2$. Fig. 2 shows the x-ray diffraction (XRD) of the deposited film after rapid thermal annealing (700°C) for 30 seconds. The inset, showing high-resolution transmission electron micrograph (HR-TEM), depicts well-defined grain boundaries in the FE film. Fig. 3 shows the P-V curve of fabricated capacitors, where the leakage is negligible.

With the same FE film formation process, we have fabricated FE-FinFETs using self-aligned gate first process on silicon-on-insulator (SOI) [7]. Fig. 4 shows the detailed fabrication process flow and schematic illustration. Fig. 5 shows the TEM cross-section of a fabricated FE-FinFET with 18 nm top fin width and 5.8 nm thick FE layer. The $I_d$-$V_{gs}$ curve shows counter-clockwise indicating FE switching (Fig. 6).

## Adaptive Program and Read Scheme

For gate length independent fast and low power programming with high fidelity, we applied 100ns write pulse to program and erase the FE-FinFETs. Fig. 7(a) shows the resultant high infidelity in the channel $G_{ch}$ states for deeply scaled devices, when the devices are programmed with a single pulse. A novel adaptive program and erase algorithm overcomes such limitation (Fig. 7(b)). The stability of read gate voltage ($V_{g,read}$) is of utmost important when implementing this algorithm. $V_{g,read}$ should not alter the programmed state and should be free from flicker noise. Therefore, a sanity check of $V_{g,read}$ induced $G_{ch}$ is conducted in the beginning. After we find a stable range of $V_{g,read}$, the devices are programmed to the lowest resistance state (11) using a 100ns pulse at 4.5V and a read operation is performed. If, the $G_{ch}$ is found to be within the desired range, the device is erased to highest resistance state (00); otherwise we alter $V_{g,read}$ to obtain the desired $G_{ch}$. Similar operation is conducted for the states "01" and "10". Fig. 7(c) shows the resultant states obtained through such closed-loop adaptive program and read scheme. Fig. 8 shows proposed architecture for implementing this algorithm at the array level. Array-level inference throughput optimization for the proposed scheme requires further system level study.

Random Gaussian distribution $G_{ch}$ is formed with measured mean and variance after closed-loop write operation to evaluate the variation for each state (Fig. 9). Fig. 10 shows DNN training process for MNIST digits recognition for both binary and 4-level/cell (multi-level cell, MLC) synapse, achieving maximum accuracy of 95.5% and 97.5%, respectively. The impact of device to device (D2D) variation that arise from inaccurately programmed $G_{ch}$ shows a mere 0.3% degradation for the MLC and 0.06% for binary cell (Fig. 11).

## IMPACT OF TEMPERATURE CHANGE

Although the adaptive program and read scheme minimizes the impact of process variation, temperature change in real scenario causes mobility and carrier concentration fluctuation, leading to change in $G_{ch}$ and $V_{th}$ for FE-FinFET synaptic devices, thus altering the stored weights in DNN. To fathom this effect, temperature dependence of device is characterized by first programming it to a fixed (low or high resistance) state at room temperature (300ºK), measuring $I_d$-$V_{gs}$ with a non-destructive $V_{gs}$, reducing temperature and measuring again. The evolution of $V_{th}$ with change of temperature for the low-resistance state (LRS, programmed using a +4.5V pulse) has been captured (Fig. 13). A FE-FET compact model that is continuous and valid from sub-threshold to strong inversion is adopted [8], along with BSIM temperature scaling expressions [9], to model $I_d$-$V_{gs}$ (Fig. 13). The same expressions are used for the high-resistance state (HRS) as well, also with good agreement between model and measured data. Fig. 14 shows the measured and modeled change in $G_{ch}$ with temperature reduction, which highlights the dependence of ON/OFF ratio on $V_{g,read}$. It is evident that, to ensure high ON/OFF ratio the device should be in subthreshold region at all temperatures in HRS, which is key to proper operation of binarized neural networks (BNN). Extension of temperatures above 300°K may be more challenging [5] and related research is in progress.

We further applied the calibrated model to consider a realistic scenario (Fig. 15), where a DNN was trained at room temperature and driven to lower temperature to perform inference without re-calibration. A BNN is considered where the devices are programmed to either HRS or LRS only. According to Fig. 14, $V_{g,read}$ of 0.75V ensures requisite ON/OFF ratio (HRS stays in the sub-threshold region). The inference accuracy remains unfazed amidst temperature change only when $V_{g,read}$ is 0.75V. For all other cases, significant accuracy drop is noted with at least one state switching from strong inversion to sub-threshold. $V_{g,read}$ optimization is necessary to ensure the preservation of the digital nature (high ON/OFF ratio) of DNN weights. Modeling of $G_{ch}$ from sub-threshold to strong inversion is crucial for such conclusion. Furthermore, we recorded retention up to $10^4$ seconds and extrapolated retention over 10 years at all temperatures (Fig. 17).

## IMPACT OF FLICKER NOISE

The flicker noise at various bias conditions is shown in Fig. 18(a). The current fluctuation is estimated based on time-domain current fluctuation ($\Delta I_d$) (Fig. 18(b)). Across many devices, $\Delta I_d / I_d$ was found to be below 1% with devices biased in linear region. The impact on DNN inference accuracy is minimal, as we will show later.

## DEVICE FIDELITY TEST

Endurance is crucial for memory applications, and even more so for DNN. In fact, for online training [10] $G_{ch}$ for synaptic devices may be adjusted up to millions of cycles for large networks. Even for offline trained DNN, weights must be re-calibrated over time. An endurance over $10^8$ cycles, limited by measurement setup (Fig. 19), shows that the FE-FinFETs fabricated in this study may be used for offline or online neural network training. Symmetric-linear multiple levels in the synaptic devices are beneficial for online training of DNN (Fig. 20). Fig. 21 and Table I compare the linearity and symmetry of our fabricated devices with other technologies. Low cycle to cycle (C2C) variation (Fig. 22) during program-erase and read operations show stable memory operations. Finally, we evaluate the cumulative impacts of D2D variation, device aging (assuming $G_{ch}$ drops to 22.6% for HRS and 74.3% for LRS, with intermediate states linear interpolated), temperature reduction, flicker noise, C2C read variation are all added together (Table II). Scenario A is all-digital BNN case, where $V_{g,read}$ of 0.75V keeps variation-induced accuracy degradation minimal (same conclusion as Fig. 16). Scenario B is closed-loop programmed 4-level states (as in Fig. 9). Significant accuracy degradation is observed due to temperature change as states switch from strong inversion to sub-threshold. Scenario C is the case where we have ensured the FE-FinFETs remain in the on-state over entire temperature range across 4-levels. Serious degradation of DNN inference accuracy with temperature reduction is observed again. The impacts of other effects remain mild. Finally, we evaluate the cumulative impacts of D2D variation, device aging (assuming $G_{ch}$ drops to 22.6% for HRS and 74.3% for LRS, with intermediate states linearly interpolated), temperature reduction, flicker noise, C2C read variation (Table II). Scenario A is the case for BNN, where $V_{g,read}$ of 0.75V keeps all non-ideal-effects-induced accuracy degradation minimal (same conclusion as Fig. 16). Scenario B is where 4-level states are closed-loop programmed (as in Fig. 9). Significant accuracy degradation is observed due to temperature change. Scenario C is the case where we have ensured the FE-FinFETs remain in the on-state over the entire temperature range across 4-levels. Serious degradation of inference accuracy is observed with device aging induced DNN weight alteration. Separate testing (not shown in Table II) indicates that temperature effects alone (without device aging) degrades inference accuracy to 10% as well. For scenarios B and C, periodic DNN calibration at system level is required, therefore endurance requirement is higher.

## CONCLUSION

We have fabricated, characterized and evaluated the performance of deeply scaled FE-FinFETs for neuromorphic computing. We conclude that the digital nature of programmed states is key to robust DNN inference in the presence of systematic variation, whereas random variation sources have less impact. With superior endurance (>$10^8$), CMOS compatibility, good retention up to 10 years, high linearity and fast (<100ns) write speed, FE-FET is promising for next-generation in-memory computing applications.

## ACKNOWLEDGMENTS

This work was jointly supported by the Ministry of Science and Technology under grant number MOST-108-2634-F-006-008, 109-2628-E-492-001-MY3 and 108-2622-8-002-016. We are grateful to the Taiwan Semiconductor Research Institute for nanofabrication facilities and services and National Center for High-Performance Computing for GPU computing facilities.

## Introduction

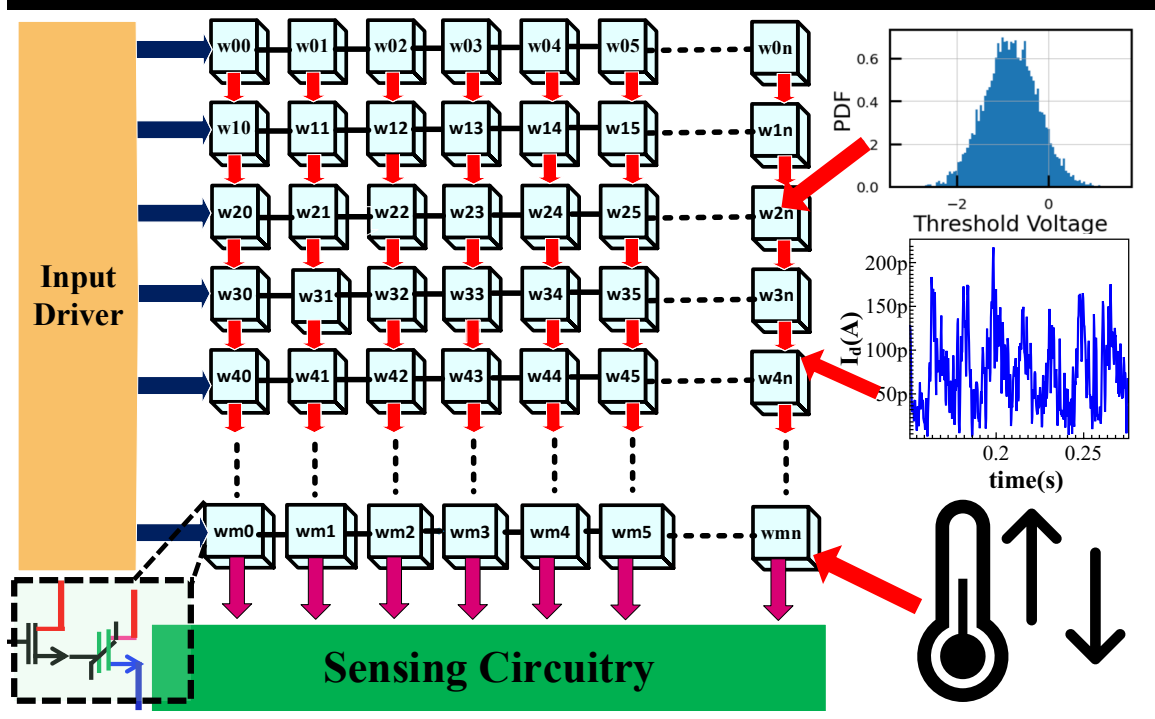

Fig. 1. Impact of Process-Voltage-Temperature on ferroelectric FET based computing-in-memory system [1].

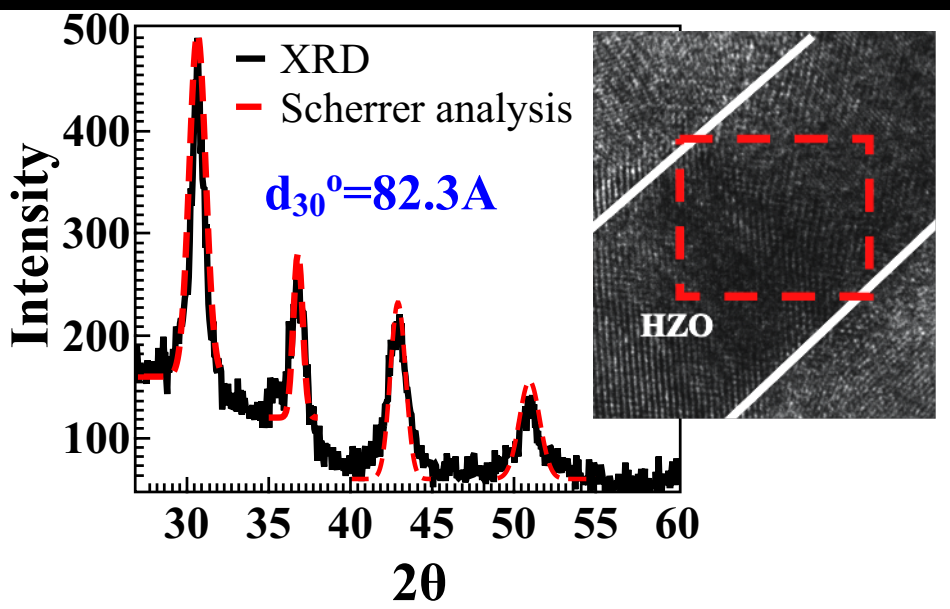

Fig. 2. XRD image of HZO film after annealing it for 30s at 700°C. The Scherrer analysis of XRD data shows the grain size is 82.3A° for the peak around 30°. The HR-TEM shows the grain boundary distribution.

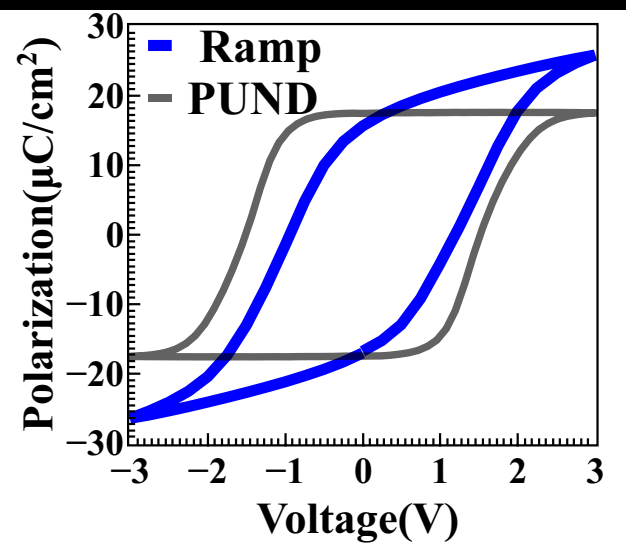

Fig. 3. The Polarization-Voltage curve(blue) shows the coercive field is around ± 1MV/cm. The black curve shows P-V curve obtained by PUND measurement.

## Fabrication and Characterization of HZO Based Ferroelectric Field Effect Transistors(FinFET)

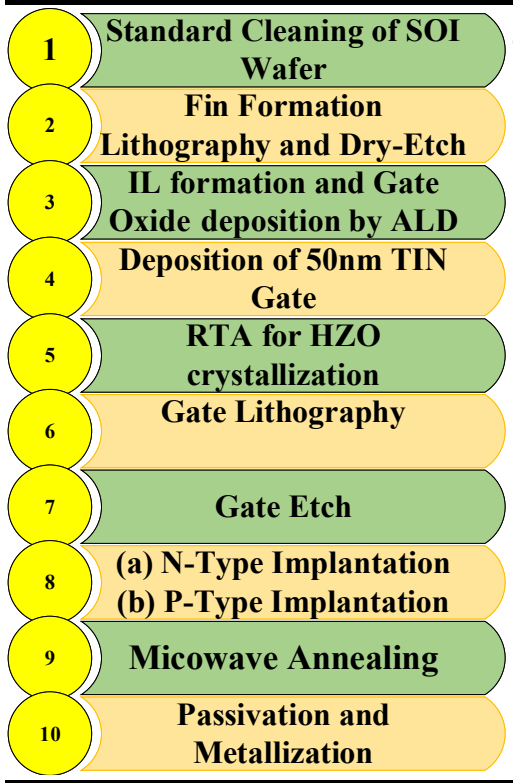

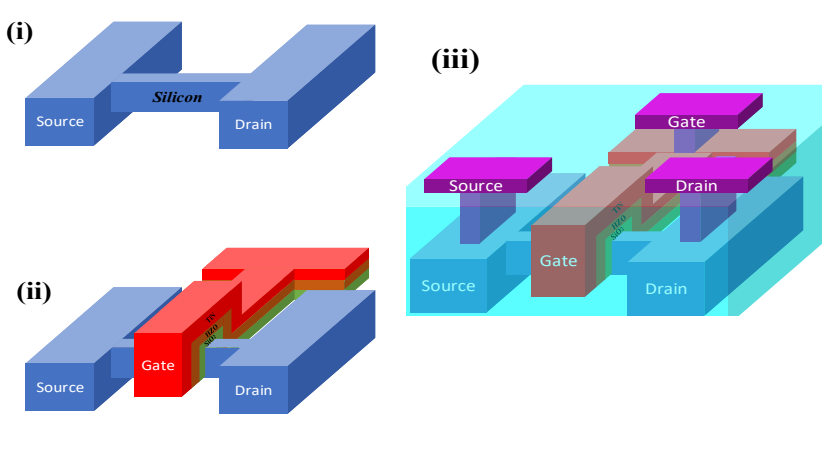

Fig. 4. Process flow for fabricating FE-FinFET device with 5nm HZO. The device schematic after (i) fin etching, (ii) post gate deposition and lithography. And (iii) post passivation and metallization.

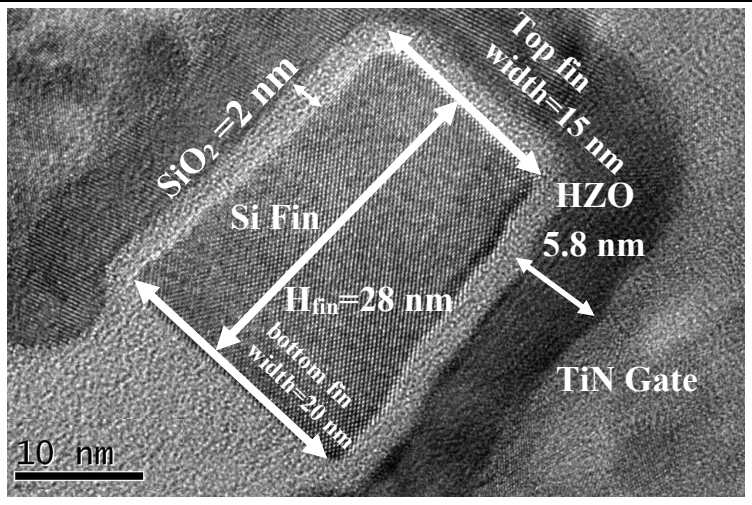

Fig. 5. TEM cross section of fabricated device. The fin width is 18nm and height is 28nm. The thickness of HZO is 5.8nm.

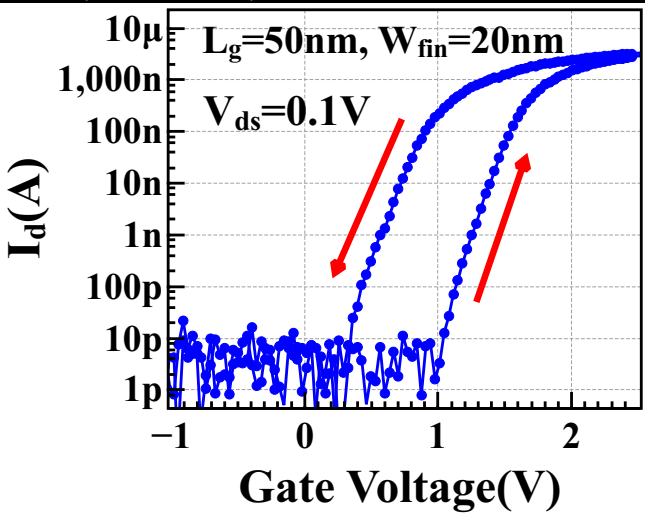

Fig. 6. The CCW swing is obtained by slowly varying gate voltage, keeping the drain at constant 100mV.

## System Level Mitigation of Process Variation

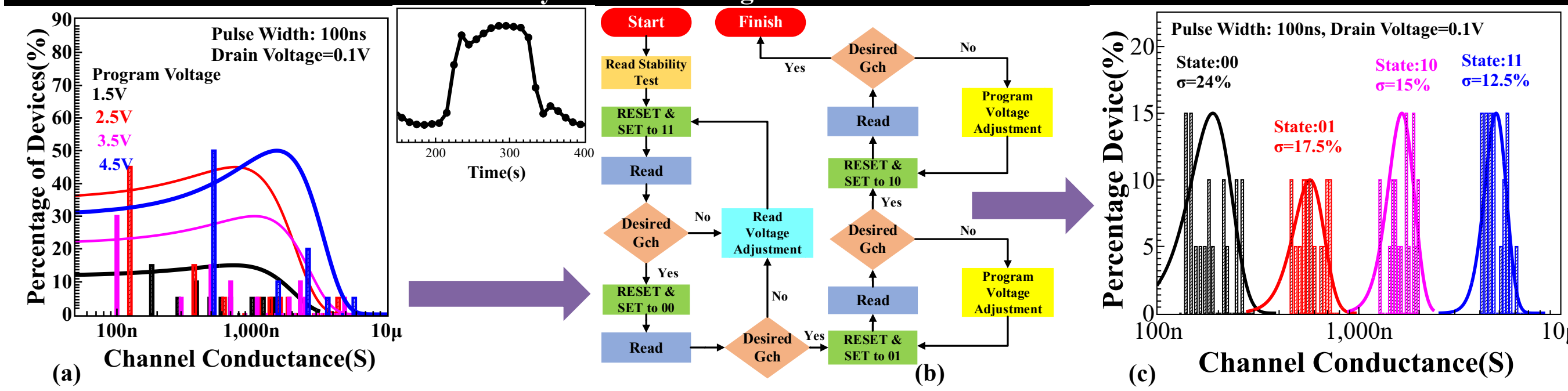

Fig.7. (a).The overlapping channel conductance states show that process-variation prevents fast programming and erase operation in deeply scaled devices. (b). System level adaptive-program-read operations algorithm deployed in this work for abating the D-to-D variation and facilitate multi-level coding in sub-10nm HZO based devices. (c). The 4-level-coding in 5nm HZO based Fe-FET.

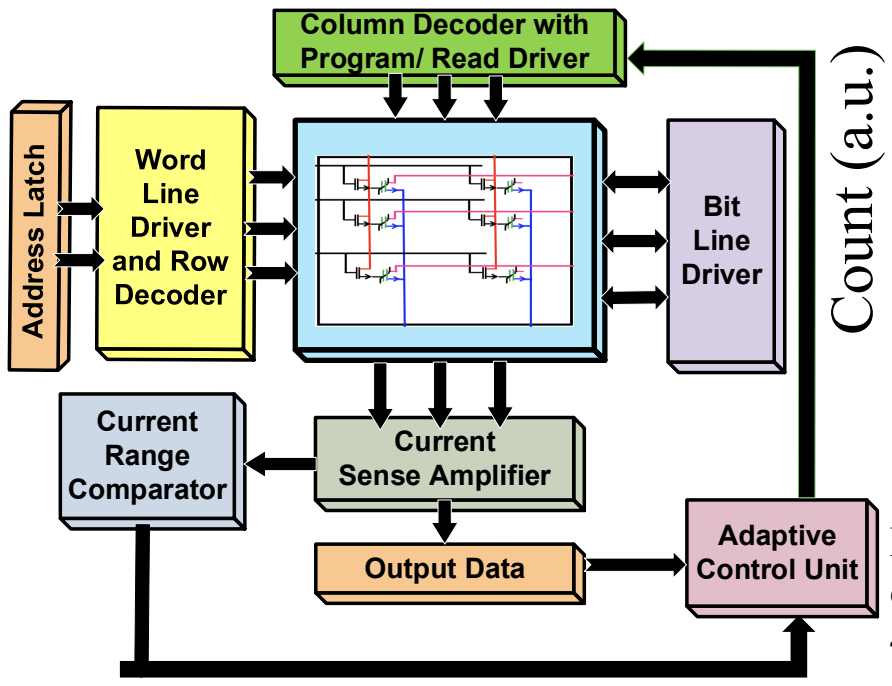

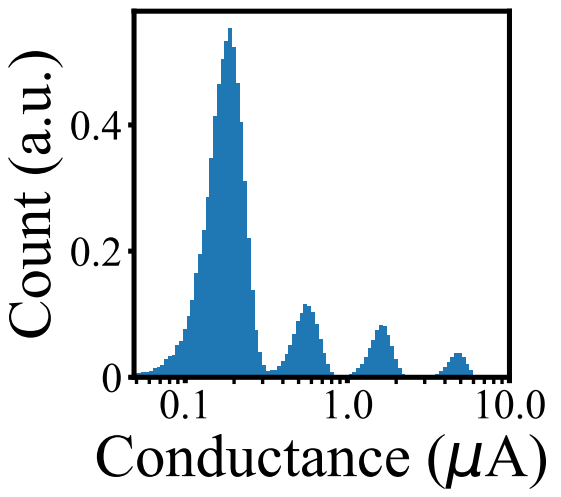

Fig. 9. The histogram for depicting the variation in 4-level multi-level cell operations.

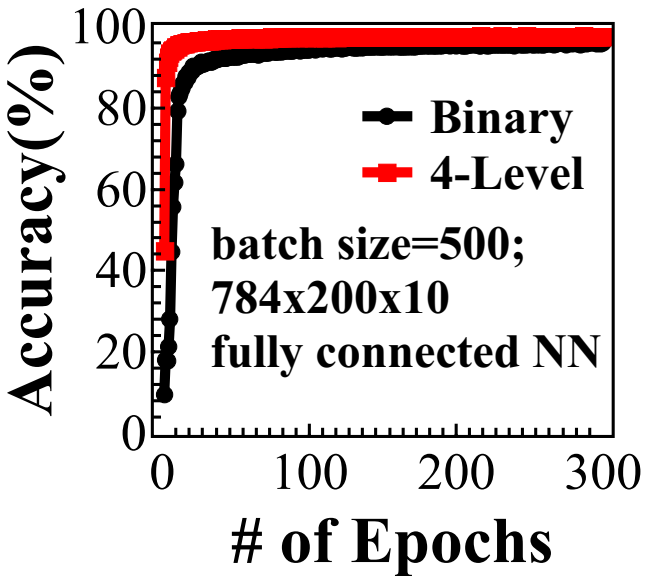

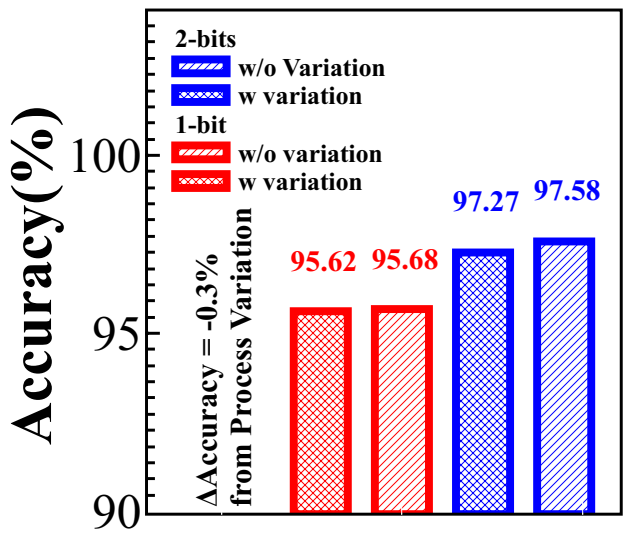

Fig. 8. Proposed block-diagram for implementing adaptive-program-read operations. (b). The histogram for depicting the variation in 4-level multi-level cell operations.

Fig. 10. Accuracy vs # of Epochs for binary and 4-level operations. The simulation is conducted without considering device to device variation present. The 4-level cell shows better performance then binary cell.

Fig. 11. The degradation in MNIST data recognition accuracy is only 0.3% in presence of device to device variation. Device to device variation is insignificant in CIM operations.

## Impact of Temperature Variation

$UA(T) = UA(300K)\left(\frac{T}{TNOM}\right)^{UA1}$

$V_{tlin}(T) = V_{tlin}(300K) + KT1.\left(\frac{T}{TNOM}-1\right)$

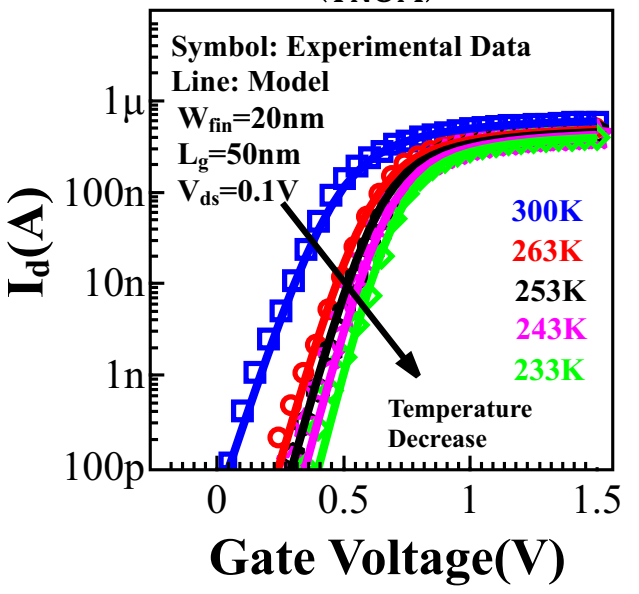

Fig. 12. Characterization and modeling the impact of temperature change for the low resistance state.

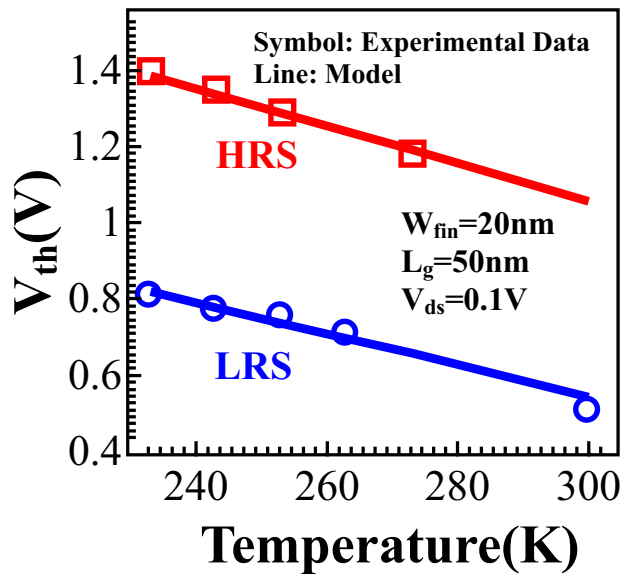

Fig. 13. Characterization and modeling the temperature change induced threshold voltage shift in FE-FinFET.

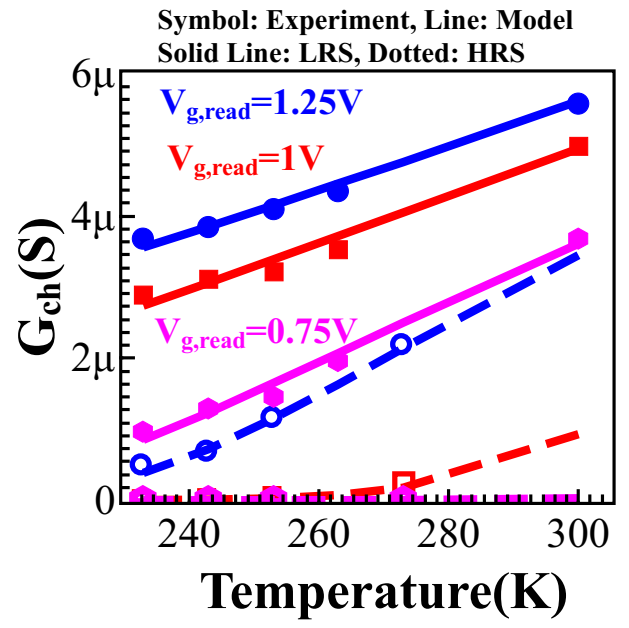

Fig. 14. Characterization and modeling the conductance drift as a result of temperature change for different read voltage.

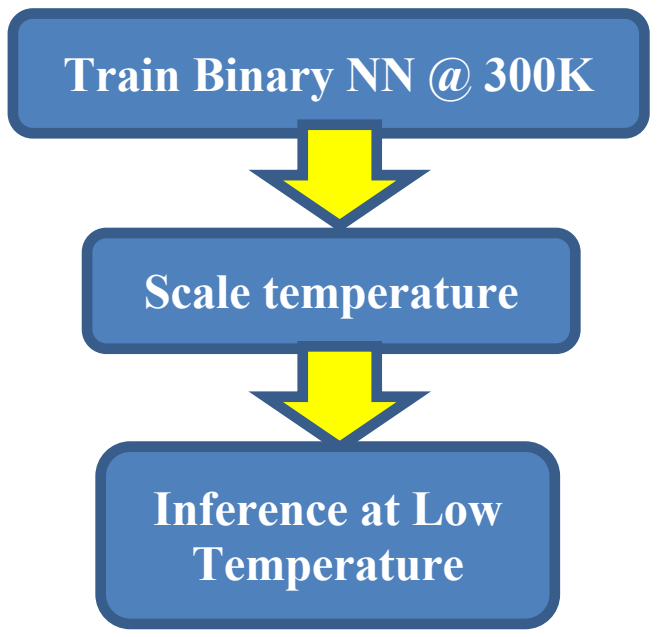

Fig. 15. Flow-chart of neuromorphic simulation for obtaining inference at lower temperature after training at room temperature.

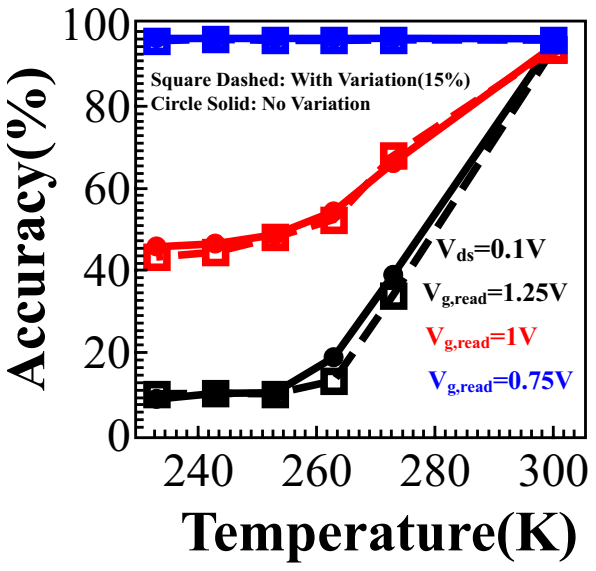

Fig. 16. Optimal choice of read voltage is a requisite to avoid accuracy degradation due to temperature induced conductance drift.

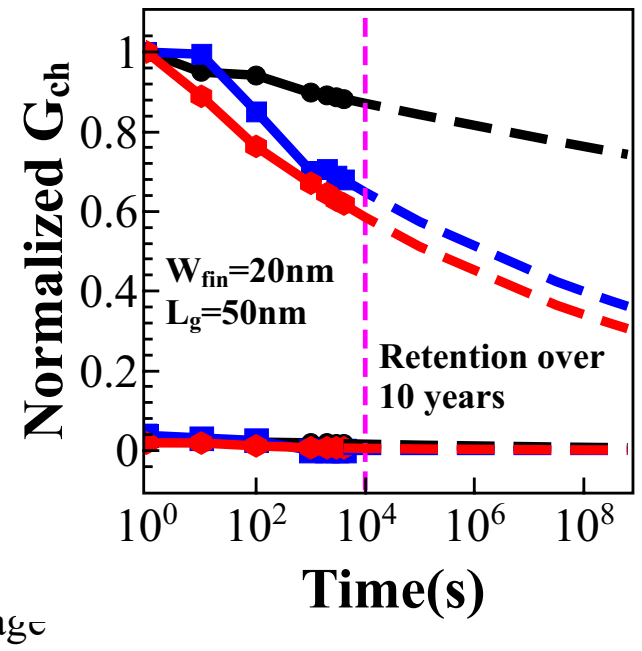

Fig. 17. Linear extrapolation shows data retention over 10 years at three different temperature.

## Impact of Flicker Noise

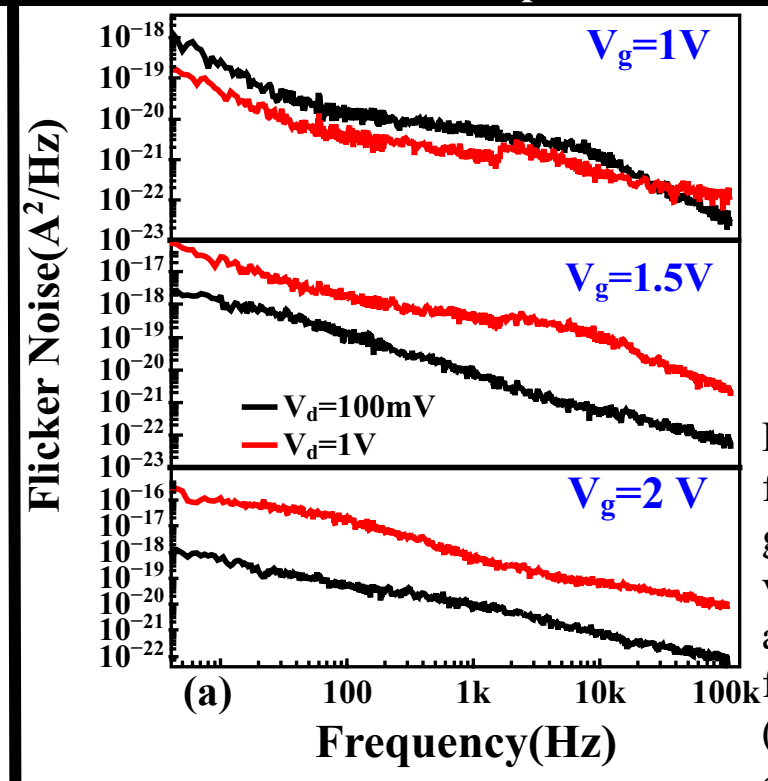

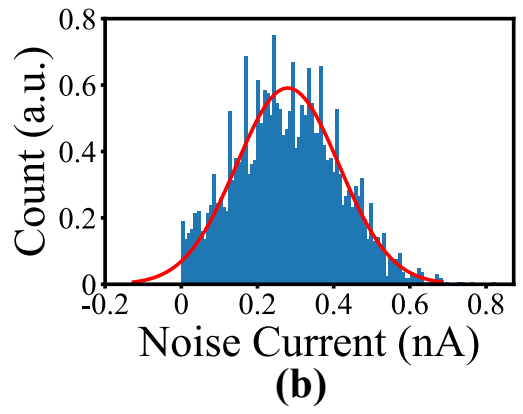

Fig. 18. (a). The flicker noise follows Lorentz model at low gate voltage and as the gate voltage is increased the trapping and detrapping increases and the flicker noise follows 1/f trend. (b). The distribution of noise current

## Device Fidelity Test

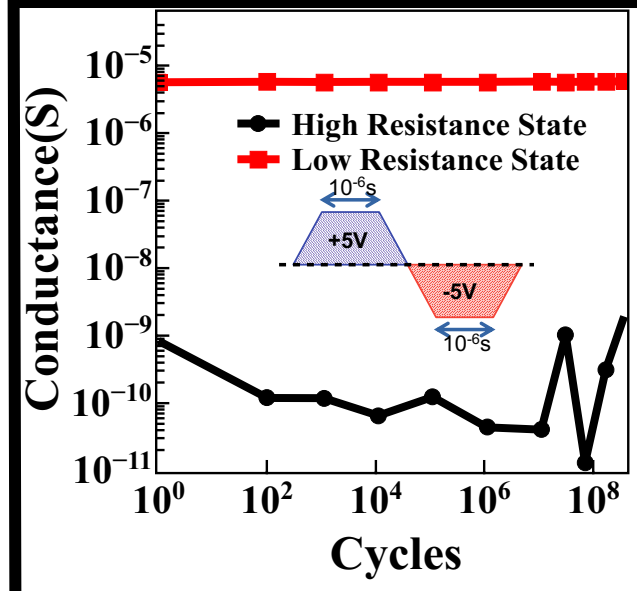

Fig. 19. Endurance up-to $10^8$ cycles shows the capability of the device being used for offline training of neural networks.

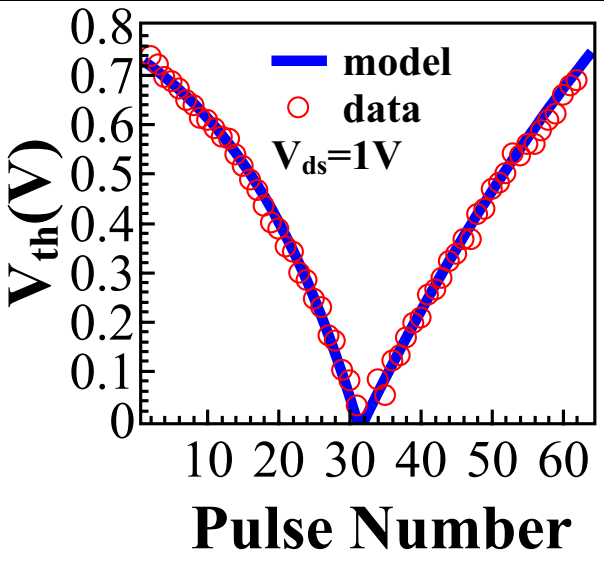

Fig. 20. Modelling the gradual change of threshold voltage during analog program and erase operations.

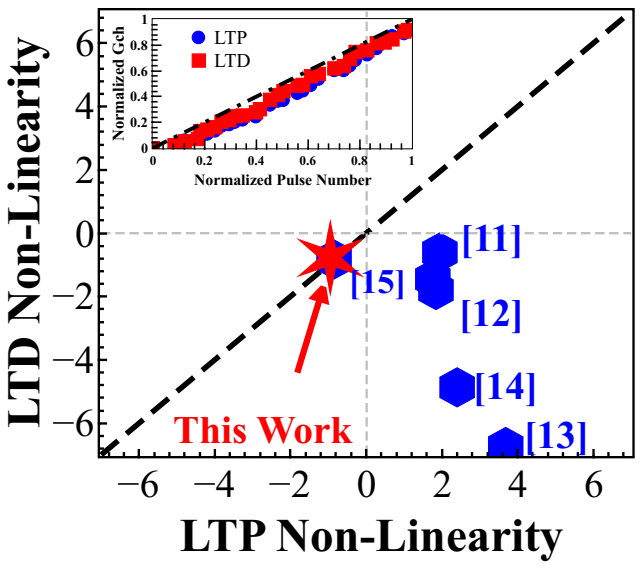

Fig. 21. Highly linear and symmetric LTP LTD characteristics in the fabricated FE-FinFET devices.

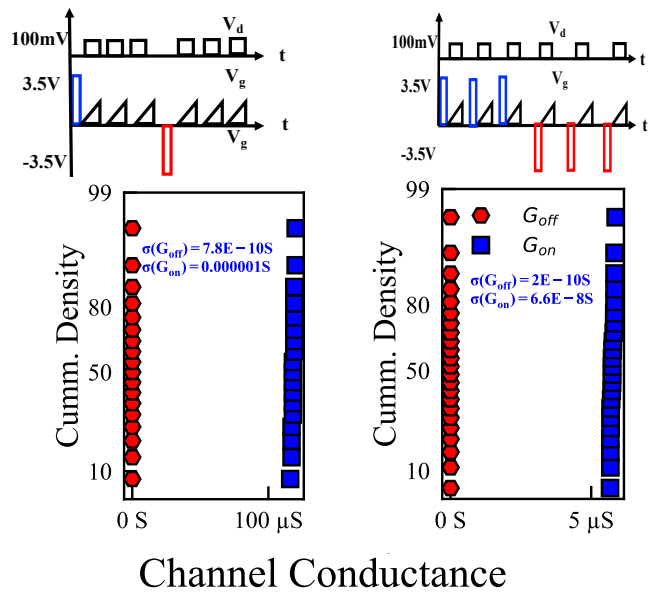

Fig. 22. Highly stable (a) read (b) program-erase operation in a device with low cycle to cycle

## Device Level Benchmarking:Table-I

| Attributes | | This Work | [4] | [16] | [17] |
|---|---|---|---|---|---|
| Cell Area(μm²) | | $10^{-3}$ | 0.25 | 0.025 | 0.026 |
| FE Thickness(nm) | | 6 | 20 | N/A | 15 |
| Program Pulse | Voltage | ±4.5 | ±10 | ±4.2 | ±10 |
| | Width | 100ns | 300ns | 10ns | 100ns |
| Endurance | | $>10^8$ | $10^4$ | $10^4$ | $10^4$ |

## Variation Impact on DNN Inference:Table-II

| | Scenario A Digital/ (Fig. 16, $V_{g,read}$=0.75] | Scenario B MLC (Fig. 9) | Scenario C (MLC analog weights) |
|---|---|---|---|
| Baseline (300°K) | 96.4% | 97.6% | 95.9% |
| Device-to-device $G_{ch}$ Variation | 96.1% [$\Delta G_{ch}$=15%] | 97.3% [$\Delta G_{ch}$ meas.] | 94.7% [$\Delta G_{ch}$=15%] |
| Device Aging $G_{ch}(t)$=-23%~-74% | 96.1% | 95.9% | 10.0% |
| Temperature 233°K | 96.1% | 46.6% | 10.0% |
| Flicker Noise ($\sigma I_d$ = 0.7%) | 96.0% | 46.3% | 10.0% |
| C2C Variation ($\sigma I_d$ = 1.2%) | 96.0% | 45.8% | 10.0% |